# On the Rational Relationship between Heisenberg Principle and General Relativity

Xiao Jianhua

Henan Polytechnic University, Jiaozuo City, Henan, P.R.C. 454000

**Abstract:** The research shows that the Heisenberg principle is the logic results of general relativity principle. If inertia coordinator system is used, the general relativity will logically be derived from the Heisenberg principle. The intrinsic relation between the quantum mechanics and general relativity is broken by introducing pure-imaginary time to explain the Lorentz transformation. Therefore, this research shows a way to establish an unified field theory of physics.

**1. Introduction**

There are many documents to treat the relationship between the Newton mechanics and general relativity. There are also many documents to treat the relationship between the quantum mechanics and general relativity. For some researchers, the related research shows that an unified field theory be possible. For others, the actual efforts to establish an unified field theory for physics have shown that such an unified field theory is totally non-possible. The physics world is divided into three kingdoms.

Hence, the title of this paper may make the reader think: See! Another nut! However, the unification of matter motion is the basic relieves for science researchers. The real problem is to find the way. This is the topic of this paper. The paper will firstly derive the Heisenberg principle from general relativity principle. Secondly, the general relativity principle is derived from Heisenberg principle. Finely, the way to construct an unified field theory is suggested.

**2. Derive Heisenberg Principle from General Relativity**

Based on the general idea of Einstein space-time theory, the vacuum space-time is described by the four-dimensional co-moving coordinator system is defined by anti-covariant coordinators ($x^1, x^2, x^3, x^4$) and initial base vectors ($\vec{g}_1^0, \vec{g}_2^0, \vec{g}_3^0, \vec{g}_4^0$). Where, $x^4$ is taken as time coordinator. From physical consideration, the initial co-moving coordinator system can be defined as the standard three-dimensional space adding time-dimension, in which following conditions are met:

$$g_{ij}^0 = \vec{g}_i^0 \cdot \vec{g}_j^0 = \begin{cases} 0, i \neq j \\ 1, i = j \neq 4 \\ -c^2, i = j = 4 \end{cases} \qquad (1)$$

Where, $c$ is light speed in "vacuum".

In general relativity, matter (motion) can be expressed by displacement field $u^i$ measured in standard physical measuring system. For any matter, the matter in discussing has invariant space coordinators, but its local base vectors are changed into ($\vec{g}_1, \vec{g}_2, \vec{g}_3, \vec{g}_4$). According to Chen Zhida's research, for incremental motion of matter, there are relationship between initial base vectors and current base vectors:

$$\vec{g}_i = F_i^j \vec{g}_j^0 \tag{2}$$

Where, the transformation tensor $F_j^i$ is determined by equation:

$$F_i^j = u^j\big|_i + \delta_i^j \tag{3}$$

where, $\big|_i$ represents covariant differentiation, $\delta_i^j$ is unit tensor.

It is clear that the incremental motion of matter is viewed as "deformation" of space-time geometry. So, gravitation field is viewed as matter "deformation" respect with idea vacuum matter. According to Einstein's idea, the Newton's inertia coordinator system can be replaced by the gradient field of infinitesimal displacement. The Equation (2) shows that the displacement gradient determines the current local geometry. Hence, the transformation tensor $F_j^i$ can be named as matter motion transformation. Matter motion in four-dimensional form is:

$$\begin{vmatrix}\vec{g}_1\\\vec{g}_2\\\vec{g}_3\\\vec{g}_4\end{vmatrix} = \begin{vmatrix} 1+u^1\big|_1 & u^2\big|_1 & u^3\big|_1 & u^4\big|_1 \\ u^1\big|_2 & 1+u^2\big|_2 & u^3\big|_2 & u^4\big|_2 \\ u^1\big|_3 & u^2\big|_3 & 1+u^3\big|_3 & u^4\big|_3 \\ u^1\big|_4 & u^2\big|_4 & u^3\big|_4 & 1+u^4\big|_4 \end{vmatrix} \cdot \begin{vmatrix}\vec{g}_1^0\\\vec{g}_2^0\\\vec{g}_3^0\\\vec{g}_4^0\end{vmatrix} \tag{4}$$

Consider such a special motion: for measuring time duration $u^4$, $u^1 = u^2 = u^3 = 0$. The Equation (4) becomes:

$$\begin{vmatrix}\vec{g}_1\\\vec{g}_2\\\vec{g}_3\\\vec{g}_4\end{vmatrix} = \begin{vmatrix} 1 & 0 & 0 & u^4\big|_1 \\ 0 & 1 & 0 & u^4\big|_2 \\ 0 & 0 & 1 & u^4\big|_3 \\ 0 & 0 & 0 & 1+u^4\big|_4 \end{vmatrix} \cdot \begin{vmatrix}\vec{g}_1^0\\\vec{g}_2^0\\\vec{g}_3^0\\\vec{g}_4^0\end{vmatrix} \tag{5}$$

Its current local time gauge is:

$$g_{44} = (1+u^4\big|_4)^2 c^2 = (1+u^4\big|_4)^2 g_{44}^0 = g_{44}^0 + [2u^4\big|_4 + (u^4\big|_4)^2] g_{44}^0 \tag{6}$$

Its current local space gauge is:

$$g_{11} = g_{11}^0 + (u^4\big|_1)^2 c^2 = g_{11}^0 + (u^4\big|_1)^2 g_{44}^0 \tag{7-1}$$

$$g_{22} = g_{22}^0 + (u^4\big|_2)^2 c^2 = g_{22}^0 + (u^4\big|_2)^2 g_{44}^0 \tag{7-2}$$

$$g_{33} = g_{22}^0 + (u^4\big|_3)^2 c^2 = g_{33}^0 + (u^4\big|_3)^2 g_{44}^0 \tag{7-3}$$

$$g_{12} = (u^4|_1)(u^4|_2)c^2 = (u^4|_1)(u^4|_2)g^0_{44} \qquad (7\text{-}4)$$

$$g_{23} = (u^4|_2)(u^4|_3)c^2 = (u^4|_2)(u^4|_3)g^0_{44} \qquad (7\text{-}5)$$

Based on Schwarzschild solution, a point mass gravity field will require that for some $i = 1,2,3$, $\frac{\partial}{\partial x^i}(1+u^4|_4)^2 \neq 0$. Hence, it is reasonable to infer that some $i = 1,2,3$, $\frac{\partial u^4}{\partial x^i} \neq 0$. On this sense, a mass can exist only when:

$$\frac{\partial u^4}{\partial x^i} \neq 0, \quad i = 4, \text{ and for some } i = 1,2,3 \qquad (8)$$

It means that the mass existence is represented by a non-zero time duration (displacement) gradient.

The geometric invariant quantity for the time duration $u^4$ in current and initial configuration is, respectively:

$$(ds)^2 = g_{ij}dx^i dx^j + g_{44}(u^4)^2, \quad i,j = 1,2,3 \qquad (9\text{-}1)$$

$$(ds_0)^2 = g^0_{ij}dx^i dx^j + g^0_{44}(u^4)^2, \quad i,j = 1,2,3 \qquad (9\text{-}2)$$

In general relativity, they must be identical. To make them be equivalents, space displacement $\delta u^i$ must be introduced. As the space displacement is related with time duration, they should be put into Equation (9-2). By Equations (1), (6), (7), and (9), the $(ds)^2 = (ds_0)^2$ will produce equation:

$$[2u^4|_4 + (u^4|_4)^2] \cdot (u^4)^2 = (\delta S)^2 \qquad (10)$$

Where, the required space variation $\delta S$ during time displacement $u^4$ is defined as:

$$(\delta S)^2 = (u^4|_i)(u^4|_j)(\delta u^i)(\delta u^j) \qquad (11)$$

The Equation (10) can be rewritten as:

$$\frac{(\delta S)^2}{(u^4)^2} = 2u^4|_4 + (u^4|_4)^2 = K \qquad (12\text{--}1)$$

Or,

$$\frac{(u^4)^2}{(\delta S)^2} = \frac{1}{2u^4|_4 + (u^4|_4)^2} = \frac{1}{K} \qquad (12\text{-}2)$$

As the time displacement differentiation about time is matter intrinsic feature from physical consideration (science laws will not change for different measurement time), the ratio between the time displacement and space variation must be constant for a given "static" matter. This, in intrinsic sense, just is the kernel of Heisenberg Principle (although which takes different form for different purpose).

From equation (12-1), comparing with $\Delta x \cdot \Delta p \geq \frac{\hbar}{2}$, one finds:

$$\frac{\hbar}{2\Delta p} = \sqrt{K}u^4, \quad \Delta p = \frac{\hbar}{2\sqrt{K} \cdot u^4} \qquad (13\text{-}1)$$

It shows that the quantum moment is related with the time displace field. Let $u^4$ be unit, it means that the quantum constant $\hbar$ is determined by the time displacement differentiation about time. It should be cosmic constant.

From Equation (12-2), comparing with $\Delta t \cdot \Delta E \geq \frac{\hbar}{2}$, one finds:

$$\frac{\hbar}{2\Delta E} = \frac{\delta S}{\sqrt{K}}, \quad \Delta E = \frac{\hbar\sqrt{K}}{2\delta S} \tag{13-2}$$

It show that the quantum energy is related with the space displace field. Let $\delta S$ be unit, it means that the quantum constant $\hbar$ is determined by the time displacement differentiation about time, also.

For classical quantum mechanics, the matter intrinsic feature $\sqrt{K}$ is interpreted as wave number function of matter.

Hence, the time displacement differentiation about time defines quantum matter. Based on this understanding, the quantum mechanics is the science about time field. Therefore, the general relativity can be taken as the start point of quantum mechanics rather than be viewed as independent.

Summing up above reasoning, the Heisenberg Principle is derived from General Relativity. Furthermore, the definition of quantum moment and energy quantity is the logic conclusion of general relativity.

## 3. Derive General Relativity from Heisenberg Principle

To derive general relativity from Heisenberg principle, the definition of the initial co-moving coordinator system must be modified as an inertia coordinator system, that is:

$$g_{ij}^0 = \vec{g}_i^0 \cdot \vec{g}_j^0 = \begin{cases} 0, i \neq j \\ 1, i = j \neq 4 \\ c^2, i = j = 4 \end{cases} \tag{14}$$

From the Heisenberg Principle, the general form of geometrical quantity must be bigger than a quantity which is related with the matter under discussion. That is:

$$(ds_0)^2 = g_{ij}^0 dx^i dx^j + g_{44}^0 (dt)^2 = (dS)^2 + c^2 (dt)^2 = (\hbar)^2 (L_0)^2 \tag{15}$$

If the matter position is accurately measured during time interval $dt$, the equation can be rewritten as:

$$(ds_0)^2 = g_{ij}^0 dx^i dx^j + g_{44}^0 (dt)^2 = c^2 (dt)^2 = (\hbar)^2 (L_0)^2 \tag{16}$$

Observing the equations $\Delta x \cdot \Delta p \geq \frac{\hbar}{2}$ and $\Delta t \cdot \Delta E \geq \frac{\hbar}{2}$, one finds that the $L_0$ is matter feature function. It must be constant for a given "static" matter.

However, for the same matter measured in arbitrary space-time during the same time interval $dt$, the current geometrical quantity is:

$$(ds)^2 = g_{ij} dx^i dx^j + g_{44}(dt)^2 = (\hbar)^2 (L)^2 \tag{17}$$

As in the commoving coordinator system, the matter is the same matter, the following quantity must be physical feature function:

$$(ds)^2 - (ds_0)^2 = (\hbar)^2 [(L)^2 - (L_0)^2] \tag{18}$$

As the time is homogenously flowing in inertia coordinator system, according to the definition of inertia coordinator system, for the matter static in an inertia coordinator system, other inertia system must be transformed by maintaining the following quantity be invariant. That is:

$$(d\tilde{s})^2 = g_{ij} dx^i dx^j - g_{44}^0 (dt)^2 \tag{19}$$

For standard rectangular coordinator system, the invariant is:

$$(d\tilde{s})^2 = (dX)^2 + (dY)^2 + (dZ)^2 - c^2(dt)^2 \qquad (20)$$

This invariant quantity is named as Lorentz invariant. In deed, the $(d\tilde{s})^2 = (\hbar)^2[(L)^2 - (L_0)^2]$ is the matter feature function. This is observed for electromagnetic field. It is also derived by Einstein from the time synchronization for relative moving inertia coordinator system. Based on physical consideration, the matter feature is invariant for any inertia coordinator system.

Extending the Equation (20) to Riemannian geometry, Einstein establishes the basic foundation for the general relativity theory.

**4. Conclusion**

Based on this research, the intrinsic relation between the quantum mechanics and general relativity is broken by introducing pure-imaginary time to explain the Lorentz transformation. Although the development of physics shows that there is no contradict between both, the real history of physics is very roundabout. The four-dimensional inertia coordinator system (defined by Equation (14)) is ruled out from the general relativity. Some even view this as the revolution of physics from Newton to Einstein. However, based on this research, the general relativity is the logic extension of Newton mechanics.

Surely, without the Heisenberg principle, this extension is not possible. Therefore, the Heisenberg principle not only bridges up the gap between the quantum mechanics and general relativity, but also bridges up the gap between the Newton mechanics and general relativity. If the general relativity theory is modified by using real time coordinators as shown in this paper, the physics is unified. Hence, the unified field theory can be established. For a possible geometrical theory, please view [1-4].